\documentclass[showpacs,prb,twocolumn]{revtex4}

\usepackage{amsfonts}
\usepackage{graphicx}
\usepackage{amsmath}
\usepackage{amssymb}
\usepackage{latexsym}
\usepackage{psfrag}

\begin{document}
\title{Conductance quantization in graphene nanoconstrictions with mesoscopically smooth but atomically stepped boundaries}
\author{S. Ihnatsenka}
\affiliation{Department of Physics, Simon Fraser University, Burnaby, British Columbia, Canada V5A 1S6}
\author{G. Kirczenow}
\affiliation{Department of Physics, Simon Fraser University, Burnaby, British Columbia, Canada V5A 1S6}

\begin{abstract}
We present the results of million atom electronic quantum transport calculations for graphene nanoconstrictions with edges that are smooth apart from atomic scale steps. We find conductances quantized in integer multiples of $2e^2/h$ and a plateau at $ \sim 0.5 \times 2e^2/h$ as in recent experiments [Tombros {\em et al.}, Nature Physics \textbf{7}, 697 (2011)]. We demonstrate that, surprisingly, conductances quantized in integer multiples of $2e^2/h$ occur even for strongly non-adiabatic electron backscattering at the stepped edges that lowers the conductance by one or more conductance quanta below the adiabatic value. We also show that conductance plateaus near $  0.5 \times 2e^2/h$ can occur as a result of electron backscattering at stepped edges even in the absence of electron-electron interactions. 
\end{abstract}

\maketitle
   
Electrical conductances quantized in integer multiples of the fundamental quantum $2e^2/h$
are the hallmark of ballistic quantum transport in nanostructures such as semiconductor quantum point contacts,\cite{Wees88, Wharam88}
gold atomic wires,\cite{Pascual93} and carbon nanotubes.\cite{Frank98} These quantized conductances are explained theoretically in terms of the Landauer theory of transport.\cite{review2010}
However, in the case of graphene nanostructures, quantum transport calculations have shown the conductance quantization to be easily destroyed by disorder\cite{Son06, Areshkin07, Gunlycke07, Yamamoto08, Evaldsson08, Mucciolo09, Lopez09, LaMagna09, Biel09} that is ubiquitous in these systems or by abrupt bends
in the quantum wire geometry.\cite{Iyengar08} Accordingly, there have been only a few reports\cite{Lin08, Lian2010, Tombros11} of conductance quantization being observed experimentally in graphene nanostructures: Lin {\em et al.}\cite{Lin08} and Lian {\em et al.}\cite{Lian2010} demonstrated conductance quantization experimentally in graphene nanoribbons. However, the conductance steps that they observed were a few orders of magnitude smaller than the ballistic conductance quantum $2e^2/h$.  This phenomenon\cite{Lin08, Lian2010} has been explained theoretically\cite{disorder09, adsorbate11} as arising from strong electron backscattering at the edges of the electronic subbands of the ribbons due to the presence of random defects. More
 recently, Tombros {\em et al.}\cite{Tombros11} have reported the experimental observation of conductance quantization in {\em integer} multiples of $2e^2/h$, as well as a fractional conductance plateau at $\sim 0.6 \times 2e^2/h$, in a graphene nanoconstriction (GNC) at zero magnetic field. 
To minimize the effects of disorder on transport in their device Tombros {\em et al.}\cite{Tombros11} studied a short suspended GNC whose width was similar to its length and estimated to be $\sim 300$ nm. Their sample was annealed by Joule heating which resulted in the constriction being formed with curved boundaries that were smooth on the mesoscopic length scale of $\sim 100$ nm. The {\em atomic}-scale structure of the boundaries was not determined experimentally, however, the  curvature of the constriction's boundaries implies the presence of large numbers of atomic-scale steps (and possibly also other defects) along the boundaries.  In this respect the GNC of Tombros {\em et al.}\cite{Tombros11} differs  from the well known semiconductor quantum point contacts (SQPCs) \cite{Wees88, Wharam88} where the transverse electron confinement is achieved {\em electrostatically} and thus the constriction boundaries are effectively smooth on the atomic scale as well as on the much larger (submicrometer) length scale of the overall dimensions of the constriction. In the limit of extremely slow spatial variation of the confining potential, electrons are adiabatically transmitted through the SQPC or adiabatically reflected. As was pointed out by Glazman {\em et al.}\cite{Glazman} such adiabatic transport results in quantized conductances; each electronic subband that is adiabatically transmitted through the narrowest part of the SQPC at the Fermi energy contributes a quantum $2e^2/h$ to the measured conductance. If the confining potential of the SQPC varies {\em smoothly} but {\em not adiabatically}, conductance quantization may still occur, each electronic subband transmitted through the narrowest part of the SQPC at the Fermi energy again contributing a quantum $2e^2/h$ to the total conductance.\cite{Castano92}  However, to date there have been no theoretical studies of conductance quantization in constrictions with boundaries exhibiting large-scale smoothness 
but atomic-scale steps, as in the GNC of Tombros {\em et al.}\cite{Tombros11}  For this reason a definitive understanding of the conductance quantization in integer multiples of $2e^2/h$ observed by Tombros {\em et al.}\cite{Tombros11} has been lacking. Furthermore it has also been unclear whether the conductance plateau observed by Tombros {\em et al.}\cite{Tombros11} at $\sim 0.6 \times 2e^2/h$ was the result of electron-electron interactions (as is widely believed of the $0.7 \times 2e^2/h$ plateau in SQPCs\cite{Thomas96, specialissue}) or whether it can be accounted for instead by strong single-electron scattering at steps in the constriction's boundaries.  

In this paper we report the results of quantum transport calculations that address these issues. We consider a non-interacting electron tight-binding model of graphene constrictions having similar dimensions to the GNC of Tombros {\em et al.}\cite{Tombros11} and having boundaries that are smooth on the length scale of the constriction but with large numbers of steps on the atomic scale. We show that this model exhibits {\em integer and fractional} conductance plateaus similar to those that were observed experimentally.\cite{Tombros11} Our results depend qualitatively on both the width of the constriction and its orientation. For the armchair orientation, the calculated integer quantized conductances of the constrictions have {\em smaller integer} values than those of uniform armchair graphene ribbons with the same width as the narrowest part of the constriction. This differs {\em qualitatively} from the well known behavior of SQPCs where the {\em adiabatic} and {\em non-adiabatic} quantized conductance values are equal to those of a uniform quantum wire whose width equals that of the narrowest part of the constriction.\cite{Glazman, Castano92} We find plateaus with conductance values $ \sim 0.5 \times 2e^2/h$ as well as the integer plateaus. For the zigzag orientation the calculated integer quantized conductances of the constrictions are either the same as or lower than those of uniform ribbons of the same width as the constriction. We also find integer and fractional quantized conductances for constrictions whose narrowest parts are neither zigzag nor armchair.

 %*********************************************************
\begin{figure}[t]
\includegraphics[scale=1.0]{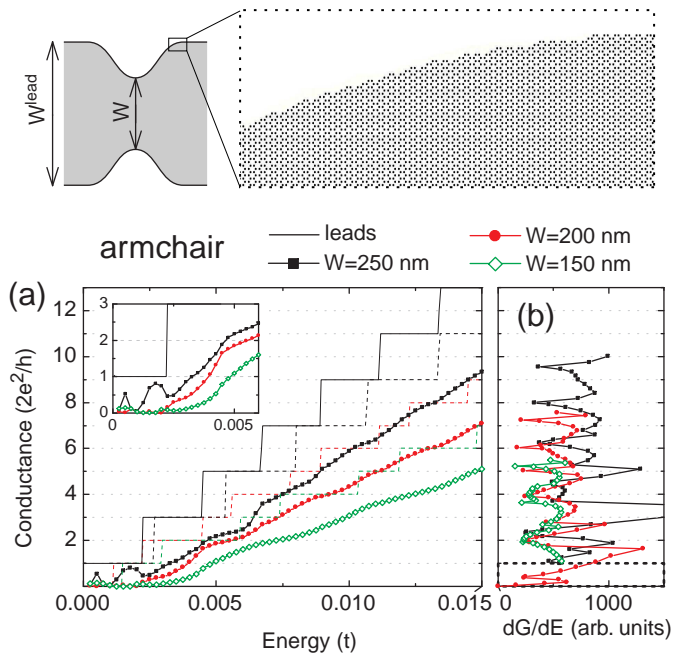}
\caption{(color online) (a) Conductance vs. Fermi energy for GNCs with constriction widths $W$=150, 200, 250 nm. The constriction shape follows the cosine function and is smooth apart from steps on the atomic scale. Host and edge orientation of the semi-infinite leads is armchair. Dotted lines show conductances of uniform armchair ribbons of the same widths as the narrowest parts of the GNC's whose conductances are plotted in the same colors. The black solid line shows the conductance of a 300 nm wide ribbon, whose width equals that of the leads $W^{lead}$. (b) Conductance $G$ vs. its energy derivative $dG/dE$. Dips in $dG/dE$ indicate conductance plateaus. In the dashed rectangle $dG/dE$ is shown for $W=200$ nm only. Temperature $T=0$. $t=2.7$ eV. The subband spacing is an order of magnitude larger than $k_BT$ even at 4.2K as in Ref.\onlinecite{Tombros11}.}
\label{fig:1}
\end{figure}
%*********************************************************
 %*********************************************************
\begin{figure}[b]
\includegraphics[scale=1.0]{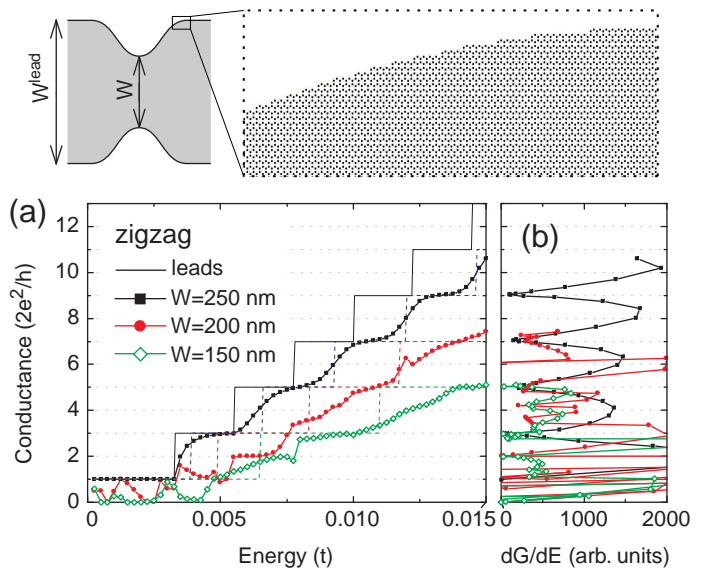}
\caption{(color online) The same as Fig. \ref{fig:1} but for the zigzag configuration of host and edges in the leads.}
\label{fig:2}
\end{figure}
%*********************************************************

 We describe GNCs by the standard tight-binding Hamiltonian on a honeycomb lattice,
\begin{equation}
 H=\sum_{i}\epsilon _{i}a_{i}^{\dag }a_{i}-\sum_{\left\langle i,j\right\rangle }t_{ij}\left( a_{i}^{\dag }a_{j}+h.c. \right),
 \label{eq:hamiltonian}
\end{equation}%
where $\epsilon _{i}$ is the on-site energy and $t_{ij}=t=2.7$ eV is the matrix element between nearest-neighbor atoms. This Hamiltonian is known to describe the $\pi$ band dispersion of graphene well at low energies.\cite{Reich02} Spin and electron interaction effects are outside of the scope of our study. The nanoconstriction and any random edge disorder and bulk 
vacancies that are present are introduced by removing carbon atoms from a uniform ribbon and setting appropriate hopping elements $t_{ij}$ to zero. It is assumed that atoms at the edges are always attached to two other carbon atoms and all dangling bonds are passivated by a neutral chemical ligand, such as hydrogen, so that the bonding between the carbon atoms at the edge and around vacancies is similar to that in bulk graphene. Random bulk and edge disorder (when present) are characterized by the probability of the carbon atoms being removed, $p^b$ and $p^e$, respectively. $p^b$ is normalized relative to the whole sample, while $p^e$ is defined relative to an edge only. The long-range potential due to charged impurities is approximated by a Gaussian form\cite{Mucciolo09, Yamamoto08} of range $d$: 
$\epsilon_i=\sum_{r_0} V_0 \text{exp}({-\left|r_i-r_0\right|^2}/{d^2})$,
where both the amplitude $V_0$ and coordinate $r_0$ are generated randomly. 

 In the linear response regime the conductance of the GNC is given by the Landauer formula\cite{review2010}
\begin{equation}
	G = \frac{2e^{2}}{h} \sum_{ji} T_{ji}.
	\label{eq:conductance}
\end{equation}
$T_{ji}$ is the transmission coefficient from subband $i$ in the left lead to the subband $j$ in the right lead, at the Fermi energy. $T_{ji}$ is calculated by the recursive Green's function method, see Ref. \onlinecite{Igor08} for details. The average conductance $\left\langle G \right\rangle$ for samples with random disorder was calculated by averaging over an ensemble of samples with different realizations of the disorder. For the results presented below, averaging was carried out over ten realization for each disorder type.

 To investigate the transport properties of GNC's we chose geometries similar to those studied experimentally.\cite{Tombros11} The shape of constriction was modeled by a cosine function so that its edges were smooth apart from atomic scale steps. The width of narrowest part of the GNC was varied in the range $W=150...250$ nm. The GNC was attached at its two ends to semiinfinite leads represented by ideal nanoribbons of width $W^{lead}$=300 nm. This guarantees that for any $W$ the leads supply more states for propagation than can pass through the narrowest part of the constriction. The region of the constriction itself in our tight-binding quantum transport calculations  included up to $\sim 1.5$ million carbon atoms. In our modeling of the effects of random disorder we assumed it to be present only in a finite region of width 300 nm and length $L=300$ nm; the semi-infinite leads were free from disorder.  
 
The calculated conductances of GNCs with different constriction widths $W$ are shown in Figure \ref{fig:1}(a) for the armchair orientation of the graphene host and edges of the ideal leads. Note, however, that the edge orientation along most of the constriction itself is neither armchair nor zigzag; see the outset in Figure \ref{fig:1}. The conductance shows faint quantization steps in integer multiples of $2e^2/h$, similar to those observed experimentally by Tombros {\em et al.}\cite{Tombros11} For better visualization we plot the energy derivative of the conductance $dG/dE$ in Fig. \ref{fig:1}(b). Here a dip in $dG/dE$ indicates a plateau in the conductance. The prominent dips in $dG/dE$ in Fig. \ref{fig:1}(b) cluster around conductance values that are integer multiples of $2e^2/h$, including both odd and even integer multiples. The conductance of the GNC decreases as constriction becomes narrower, a feature expected theoretically and observed\cite{Wees88, Wharam88} in conventional semiconductor quantum point contacts: As the constriction width shrinks the number of propagating states for a given Fermi energy decreases. Interestingly, although the conductance plateaus occur near integer multiples of $2e^2/h$, in each case the integer has a {\em smaller} value than that for the ideal infinite ribbon of uniform width
whose width equals the width $W$ of the narrowest part of the constriction, for the same electron Fermi energy, calculated with the same tight binding approach. This can be seen by comparing the conductances of the GNC's in Fig. \ref{fig:1}(a) with those of the corresponding uniform ideal ribbons\cite{review2012,otherribbons} that are shown as the dotted lines of the same color in Fig. \ref{fig:1}(a). We also found no correlation between the calculated GNC conductances and the semiconductor/metallic property of uniform ideal armchair ribbons. These findings show that the conductance quantization that we find for the armchair oriented host and leads is {\em not} due to adiabatic transmission of individual eigenmodes of the ideal leads through the constriction but that additional scattering along the constriction edges plays an important role.   

%*********************************************************
\begin{figure}[t]
\includegraphics[scale=1.0]{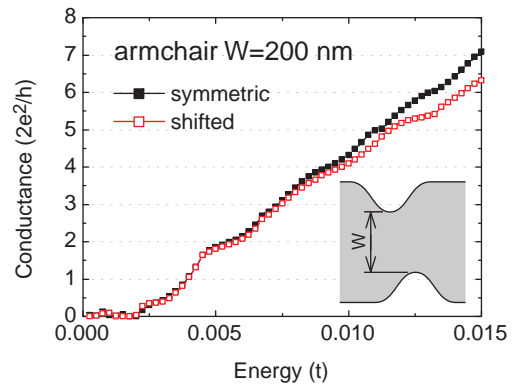}
\caption{(color online) Conductance vs. Fermi energy for GNC of width W=$200$ nm. The red open squares show the conductance for a constriction whose top and bottom parts are shifted by 80 nm relative to each other. Black filled squares show for comparison the conductance for the corresponding symmetric constriction of width $W=200$ nm, as in \ref{fig:1}(a). }
\label{fig:3}
\end{figure}
%*********************************************************
%*********************************************************
\begin{figure}[t]
\includegraphics[scale=1.0]{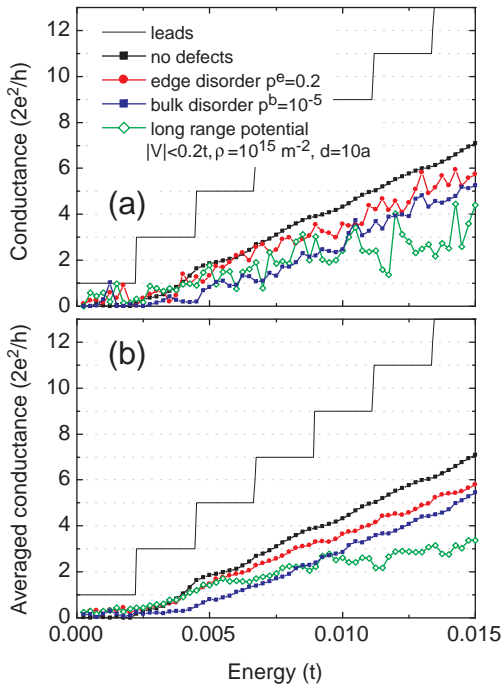}
\caption{(color online) (a) Effect of different disorder types on the conductance of armchair-oriented GNCs. (b) Conductance averaged over 10 realization of disorder. Constriction width $W$=200 nm. Red line with filled circles is for edge disorder with $p^e=0.2$. Blue line with filled squares is for bulk vacancy disorder with $p^b=10^{-5}$. Green line with rhombuses is for long ranged potentials due to charged impurities with effective parameters $|V|\leq0.2t$, $\rho=5\times10^{15}$ m$^{-2}$, $d=10a$. The black solid line shows the conductance for an ideal, uniform ribbon 300 nm wide and is given as a reference.}
\label{fig:4}
\end{figure}
%*********************************************************

For $W=200$ nm, we find an additional conductance step at $\sim0.5\times2e^2/h$; see Fig. \ref{fig:1}(b) and the inset of Fig. \ref{fig:1}(a). This agrees with the experimental findings in Ref. \onlinecite{Tombros11}. Whether or not this feature is present in the results of our quantum transport calculations depends on the width W of the constriction; note that experiments have been reported for only a single sample.\cite{Tombros11} However, as can be seen in Fig. \ref{fig:1} this fractional plateau coexists with {\em integer} conductance plateaus at higher electron Fermi energies for the same GNC and occurs for both electron and hole transport (not shown), as in the experimental data.\cite{Tombros11}

Our results for GNCs with the zigzag orientation\cite{Jia09, Engelund10} of the graphene host and edges of the leads are shown in Fig. \ref{fig:2}. Comparison of  Fig. \ref{fig:1} and  Fig. \ref{fig:2} reveals significant differences between quantized conductance plateaus in GNCs with the host and leads in the zigzag and armchair orientations: For the zigzag orientation the quantized conductance plateaus are more pronounced than for the armchair case. Also for the zigzag case the calculated values of the quantized conductances of the GNCs in many (but not all) cases are close to the values of the quantized conductances of ideal uniform zigzag ribbons having the same width as the narrowest part of the GNC and the same electron Fermi energy. By contrast, as we have already mentioned, all of the integer quantized GNC conductances for the armchair case are smaller than those of the corresponding uniform ideal ribbons by integer multiples of $2e^2/h$.   Thus in many cases non-adiabatic electron backscattering is much weaker for GNCs in the zigzag orientation than for those in the armchair orientation. This difference may be attributed to the current densities being much lower near zigzag graphene edges than near armchair edges so that the conductances are less affected by edge imperfections for zigzag ribbons.\cite{Zabro07} 

The open red squares in Fig. \ref{fig:3} show the calculated conductance of an asymmetric GNC with the armchair orientation of the host and leads. As shown in the inset, the geometry in this case is similar to the $W=200$nm armchair-oriented constriction in Fig. \ref{fig:1} except that the upper and lower regions where the carbon atoms have been removed are now offset from each other laterally by 80 nm. Thus the edges of the narrowest part of the constriction have neither the armchair nor the zigzag orientation.  We find that the electron backscattering  is somewhat stronger (the conductance lower) in this case than for the symmetric $W=200$nm armchair-oriented constriction in Fig. \ref{fig:1}; the calculated conductance for the latter is replotted as the solid black squares in Fig. \ref{fig:3} for comparison. However the first few quantized conductance plateaus (as well as the plateau at $\sim0.5\times2e^2/h$) are still clearly visible for the asymmetric GNC. 

The effects of disorder of different types are shown in Fig. \ref{fig:4}. As a test system we chose a GNC of width $W=200$ nm having the armchair orientation. The effect of disorder on the conductance of the GNC is similar to that for graphene nanoribbons.\cite{disorder09} However, the conductance quantization is strongly degraded for every disorder type including bulk vacancies. This may be attributed to the varying width of the GNC along the transport direction that precludes the existence of well-defined subband edges for the whole structure. We find each type of disorder to suppress the conductance and to result in universal conductance fluctuations.\cite{disorder09, Lee85} 

In conclusion, we have carried out million-atom electronic quantum transport calculations for graphene nanoconstrictions with boundaries that are smooth except for steps on the atomic scale and have dimensions similar to those of the graphene nanoconstrictions that have been found to exhibit conductances quantized in integer multiples of $2e^2/h$ in recent experiments.\cite{Tombros11} Our results demonstrate quantized conductances similar to those observed experimentally\cite{Tombros11} in a tight binding model with non-interacting electrons. We find conductances quantized in {\em integer} multiples of $2e^2/h$ to occur in graphene nanoconstrictions even in the presence of strong electron backscattering at the stepped constriction edges that depresses the quantized conductance values by {\em one or more} $2e^2/h$ {\em conductance quanta} below the quantized conductance values of uniform graphene ribbons with the same width and electron Fermi energy as those of the narrowest part of the constriction. This {\em integer} conductance quantization in the presence of such {\em strong} backscattering has no known analog in either adiabatic or non-adiabatic semiconductor quantum point contacts. It may explain why, based on their transport measurements, Tombros {\em et al.}\cite{Tombros11} estimated the width of their GNC to be smaller (200-275 vs. 300 nm) at zero magnetic field than at higher magnetic fields where electron backscattering at the edges of the constriction is reduced.\cite{edge_states} We also find that conductance plateaus at $ \sim 0.5 \times 2e^2/h$ need not be the result of electron-electron interactions in these systems but can result instead from non-adiabatic backscattering of electrons at atomically stepped constriction boundaries. However, the plateau observed experimentally at $ \sim 0.6 \times 2e^2/h$ by Tombros {\em et al.}\cite{Tombros11} resembles the plateau that is seen at $ \sim 0.7 \times 2e^2/h$ in SQPCs and is attributed to electron-electron interactions\cite{Thomas96,specialissue} in part because in SQPCs the potentials are smooth and there is no analog of the atomic steps present at the edges of GNCs. Therefore further experimental studies are required to clarify whether electron-electron interactions or boundary scattering are primarily responsible for the fractional plateau observed by Tombros {\em et al.}\cite{Tombros11} in the GNC. Our results (see Fig. \ref{fig:1}) suggest that systematic experimental studies of GNCs having differing widths may answer this question. Our quantum transport calculations also show random defects to strongly degrade the conductance quantization in graphene nanoconstrictions.  

%\begin{acknowledgments}
This work was supported by NSERC, CIFAR, Compute Canada and WestGrid. 
%\end{acknowledgments}

\end{document}